\documentclass[aps,showpacs,twocolumn,showkeys,prl]{revtex4}
\usepackage{graphicx,color,epsfig,rotate}
\usepackage{times,xspace}
\usepackage{amsbsy,amssymb,amsmath,bm}
\usepackage{dcolumn}
\usepackage{bm}
\input{epsf.sty}

\def\bbbc{{\mathchoice {\setbox0=\hbox{$\displaystyle\rm C$}\hbox{\hbox
to0pt{\kern0.4\wd0\vrule height0.9\ht0\hss}\box0}}
{\setbox0=\hbox{$\textstyle\rm C$}\hbox{\hbox
to0pt{\kern0.4\wd0\vrule height0.9\ht0\hss}\box0}}
{\setbox0=\hbox{$\scriptstyle\rm C$}\hbox{\hbox
to0pt{\kern0.4\wd0\vrule height0.9\ht0\hss}\box0}}
{\setbox0=\hbox{$\scriptscriptstyle\rm C$}\hbox{\hbox
to0pt{\kern0.4\wd0\vrule height0.9\ht0\hss}\box0}}}}

\begin{document}
\title{Magnetically-induced electric polarization in an organo-metallic magnet}

\author{V. S. Zapf$^1$, M. Kenzelmann$^2$, F. Wolff-Fabris$^{1,*}$, F. Balakirev$^1$, Y. Chen$^{3,4,5}$}

\affiliation{$^1$National High Magnetic Field Laboratory (NHMFL), Los Alamos National Lab (LANL), Los Alamos, NM \\
$^2$Laboratory for Developments and Methods, Paul Scherrer Institute, CH-5232 Villigen, Switzerland  \\
$^3$Department of Physics and Astronomy, Johns Hopkins University, Baltimore, Maryland 21218, USA \\
$^4$ NIST Center for Neutron Research, National Institute of Standards and Technology, Gaithersburg, MD 20899 \\
$^5$ Department of Materials Science and Engineering, University of Maryland, College Park, MD 20742\\
$^{\dagger}$ Now at Dresden Hoch-feld Labor, Dresden, Germany, D-01328}

\date{\today}

\begin{abstract}

The coupling between magnetic order and ferroelectricity has been under intense investigation in a wide range of transition-metal oxides.  The strongest coupling is obtained in so-called magnetically-induced multiferroics where ferroelectricity arises directly from magnetic order that breaks inversion symmetry. However, it has been difficult to find non-oxide based materials in which these effects occur. Here we present a study of copper dimethyl sulfoxide dichloride (CDC), an organo-metallic quantum magnet containing $S = 1/2$ Cu spins, in which electric polarization arises from non-collinear magnetic order. We show that the electric polarization can be switched in a stunning hysteretic fashion. Because the magnetic order in CDC is mediated by large organic molecules, our study shows that magnetoelectric interactions can exist in this important class of materials, opening the road to designing magnetoelectrics and multiferroics using large molecules as building blocks. Further, we demonstrate that CDC undergoes a magnetoelectric quantum phase transition where both ferroelectric and magnetic order emerge simultaneously as a function of magnetic field at very low temperatures.
\end{abstract}

\maketitle

Magnetoelectric multiferroics are compounds with magnetic and electric orders that coexist and are coupled via magnetoelectric interactions \cite{Fiebig05,Khomskii06,Eerenstein06,Hill00}. Research in this field is motivated by the promise of devices that can sense and create magnetic polarizations using electric fields and vice versa, thereby creating new functionality as well as improving the speed, energy-efficiency and size of existing circuits. A new class of induced multiferroics has become the topic of intense study in the past few years, in which a magnetic order induces an electric polarization \cite{Kimura03,Goto04,Hur04,Lawes05,Katsura05,Cheong07,Kenzelmann07,Arima07,Kimura07}. These materials are either low-dimensional or frustrated magnets in which competing interactions give rise to non-collinear order that breaks inversion symmetry. This inversion symmetry-breaking magnetism couples to the lattice most likely via spin-orbit interactions that attempt to lower magnetic entropy and in the process create electric polarization. 

Most magnetoelectrics and multiferroics are transition-metal oxides or fluorides where the magnetoelectric interactions are mediated via superexchange through the oxide and fluoride anions. Magnetoelectric interactions can also be expected in other materials such as organo-metallic solids but have not yet been clearly established \cite{Cui06,Ye08,Scott08}. In this paper we present field-induced multiferroic behaviour in an organo-metallic quantum magnet, CuCl$_2\cdot2$[(CH$_3$)$_2$SO] (CDC) where the Cu spins adopt non-collinear magnetic order that creates an electric polarization in the presence of magnetic fields. Unlike most magnetically-induced multiferroics, the magnetic spins do not form a spiral in order to break inversion symmetry.

\begin{figure}[tbp]
\epsfbox{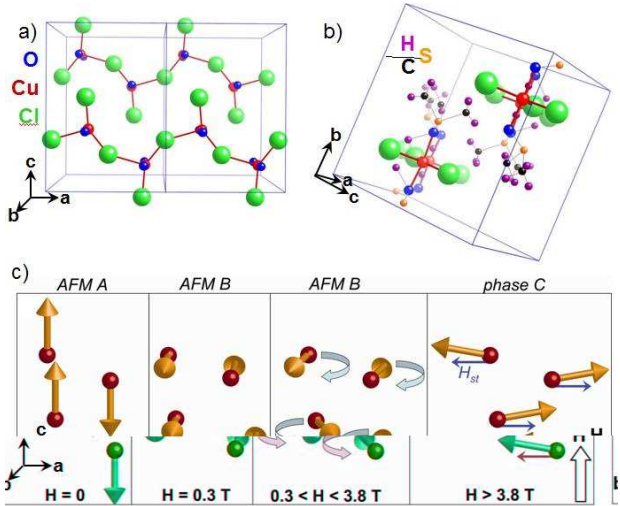}
\caption{a) Crystal structure of CDC showing the Cu-Cl chains. b) Structure showing the organic molecules mediating exchange along the b-axis. c) Effect of applied magnetic fields on the spins (shows as arrows) of the Cu atoms (shown as balls) for one unit cell.}
\label{Structure}
\end{figure}

\begin{figure}[tbp]
\epsfbox{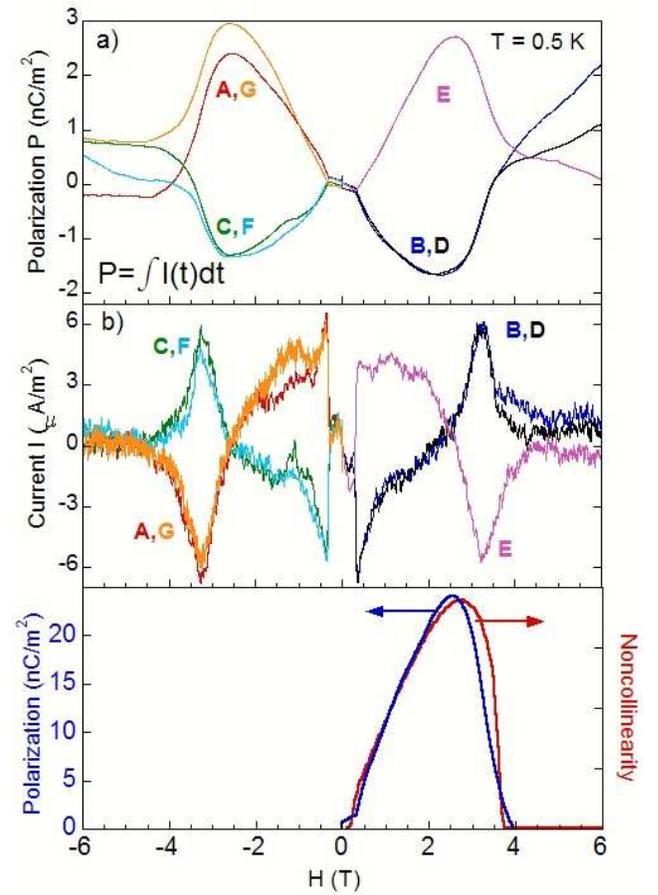}
\caption{ a) Electric polarization P along the b-axis vs magnetic field H along the c-axis collected during a sequence of seven magnetic field pulses (A-G) at 0.5 K. (Data is shown only for rising fields.) b) Magnetoelectric current I vs H. A first-order-like transition into the ferroelectric phase is evident at 0.3 T and a broader peak occurs at 3.8 T. c) Comparison of the electric polarization and the non-collinear magnetic order, which is the product of the long-range antiferromagnetic order and the field-induced staggered local fields measured by inelastic neutron diffraction [16, 17].}
\label{Polarization}
\end{figure}

CDC crystallizes in an orthorhombic crystal structure with space group Pnma (see Fig. \ref{Structure}a) \cite{Willet70}. The Cu spins form zig-zag chains in the crystallographic a-c plane, along which the Cu spins are antiferromagnetically coupled via a superexchange interaction J = 1.46 meV mediated by Cl ions \cite{Kenzelmann04}. Perpendicular to the chains, the Cu atoms are separated by dimethyl sulfoxide groups (Fig. \ref{Structure}b) and it is this weaker antiferromagnetic interaction that sets the energy scale for 3-dimensional long-range antiferromagnetic order with a N$\rm\acute{e}$el temperature T$_N = 0.9$ K \cite{Kenzelmann04}. Fig. \ref{Structure}c shows the evolution of the Cu spins with applied magnetic fields along the c-axis. At zero magnetic field $H$, the magnetic order consists of a collinear antiferromagnetic arrangement with the magnetic moments pointing along the c-axis. For magnetic fields   applied along the c-axis, the spins undergo a first-order spin-flop transition into the b-axis at 0.3 T. As $H$ is increased further, spin-orbit couplings produce a staggered g-tensor and a Dzyaloshinskii-Moriya interaction that create effective local magnetic fields on the Cu sites whose magnitude is proportional to the external magnetic field   and whose direction alternates from one site to the next. These are only shown in the final panel of Fig. \ref{Polarization}c for clarity. In response to these staggered fields, the spins gradually rotate from the b-axis to the a-axis and become non-collinear in the process \cite{Chen07}. Between $H  = 0.3$ T and 3.8 T, the magnetic structure is thus described by two antiferromagnetic order parameters, one due to the antiferromagnetic superexchange interactions between the Cu spins and the other due to the spin-orbit-induced staggered fields. Finally, the spins align with the staggered fields along the a-axis for $H_c  > 3.8$ T. In addition to the behaviour described so far, the spins also increasingly cant along the magnetic field direction with increasing field. The non-collinear state between 0.3 and 3.8 T (AFM B) where the two order parameters are competing with each other breaks inversion symmetry and thereby allows an electric polarization to occur along the b-axis. In this work we present measurements of the electric properties of CDC and we show that the magnetic and electric polarizations coexist and are closely coupled for $H_c$ between 0.3 and 3.8 T.

\epsfxsize=250pt

Single-crystal samples of CDC were grown by a flux method \cite{Chen07}. For electric polarization measurements, the samples were coated on the a-c faces with silver paint and leads were attached, permitting the magnetoelectric current to be recorded with a Stanford Research 570 current to voltage converter along the b-axis. Measurements were performed during magnetic field pulses with applied magnetic fields along the c-axis. The samples were cooled by immersion in liquid $^3$He while rapid 10 T magnetic pulses with dB/dt up to 3.8 kT/s were applied using a short-pulse capacitatively-driven magnet at the National High Magnetic Field Laboratory in Los Alamos, NM. A large $dB/dt$ increases the signal to noise levels and thus only the upsweep data with a larger $dB/dt$ is shown. The dielectric constant was measured capacitatively in the liquid mixing chamber of a $^3$He-$^4$He dilution refrigerator. In order to rule out possible magnetostriction affecting the dielectric constant measurements, magnetostriction was measured separately using a titanium dilatometer in the vacuum mixing chamber \cite{Zapf08,Schmiedeshoff06}. No signatures of the 3.8 T transition were observed with a sensitivity to magnetostriction effects 10x greater than that of the dielectric constant measurements.

\epsfxsize=250pt

The electric polarization of CDC along the b-axis and the magnetoelectric current from which it was derived are shown in Fig. \ref{Polarization}a and b. The data are shown for a series of magnetic field pulses along the positive and negative c-axis (A through G) applied at $T = 0.5$ K while the electric polarization is measured along the b-axis. The electric polarization has a dome shape for applied magnetic fields between 0.3 and 3.8 T, indicating an electric polarization that coexists with the AFM B phase. In the magnetoelectric current data of Fig. \ref{Polarization}b, the onset of electric polarization at 0.3 T is sharp and first-order like, whereas the transition near 3.8 T appears to be second order. As shown in Fig. \ref{Polarization}c, the field dependence of the electric polarization is proportional to a measure of non-collinearity, namely the product of the two-antiferromagnetic order parameters, which are determined from neutron scattering data \cite{Scott08,Willet70}. This provides firm evidence that the electric polarization is generated by non-collinear magnetic order.

\epsfxsize=250pt
\begin{figure}[tbp]
\epsfbox{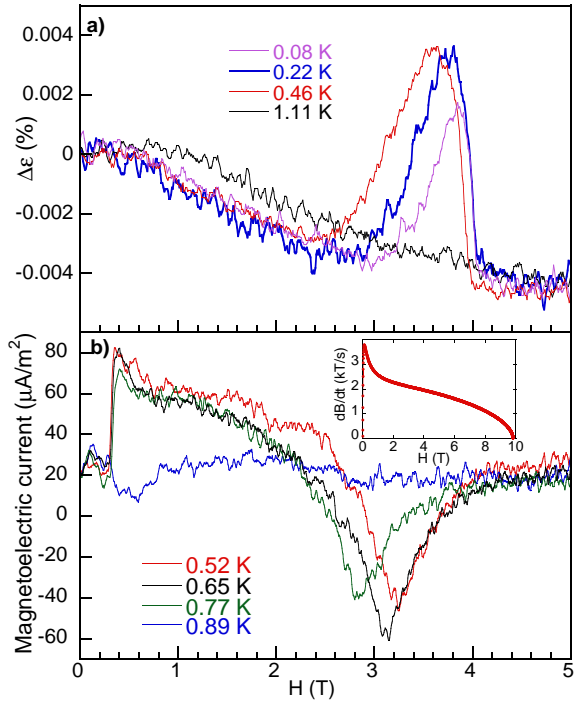}
\caption{a) Percentage change in the dielectric constant   due to applied magnetic field. b) Magnetoelectric current measurements vs. applied magnetic field showing the transitions into and out of the ferroelectric phase. Inset: Rate of change of magnetic field as a function of field during a 10 T pulse. 
}
\label{Temperature}
\end{figure}

The direction of the electric polarization can switch depending on the history of the magnetic field pulses, which demonstrates that it is ferroelectric polarization. When two consecutive magnetic field pulses are applied along the same direction in the c-axis (e.g. D followed by E or F followed by G in Fig. \ref{Polarization}a), the resulting magnetic polarization for E and G is "positive" along b. (The definition of positive and negative is arbitrary). On the other hand, when two consecutive pulses are applied in opposite directions along the c-axis (A and B, B and C, C and D, E and F), the resulting polarization for B, C, D and F is switched into the "negative" direction along the b-axis. These measurements were repeated for three samples and found to be reproducible. This type of switching behavior, where the direction of the electric polarization depends both on the direction of the present magnetic field pulse and the previous one, is unique as far as we know among magnetoelectrics. We note that in addition to the sample being intrinsically hysteretic, there must also be a symmetry-breaking electric field along the b-axis in order for the behavior described above to occur. This is further supported by the fact that no electric field poling is required to observe the electric polarization. We speculate that this could result from the influence of Schottky voltages where the capacitor plates contact the sample. These do not contribute a background signal to the data since they are magnetic field-independent, however they can subject the crystal to a symmetry-breaking electric field along b. The absence of electric-field poling required to observe electric polarization has previously been reported in the multiferroic compounds LiNiPO$_4$ and LiCoPO$_4$ \cite{Kornev00}.

\epsfxsize=230pt
\begin{figure}[tbp]
\epsfbox{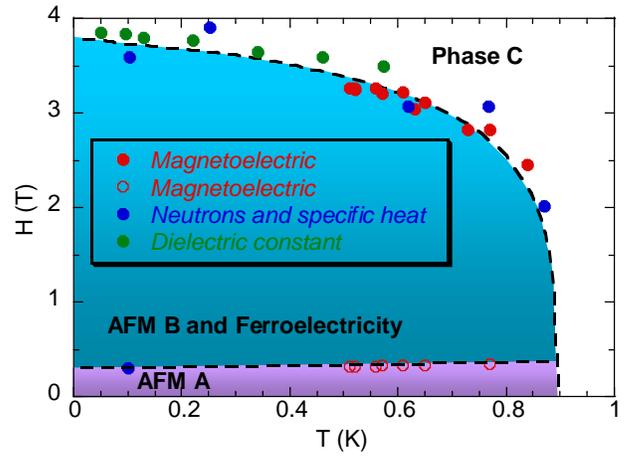}
\caption{H-T phase diagram of CDC showing regions of collinear and non-collinear antiferromagnetism (AFM A and B), ferroelectricity, and the aligned staggered paramagnetism (phase C) that occurs at high fields for H along [001]. Data are obtained from peaks in the dielectric constant and magnetoelectric current (this work), neutron diffraction [18] and specific heat [19]. 
}
\label{PhaseDiagram}
\end{figure}

The existence of a phase transition in the electric polarization is confirmed by dielectric constant measurements along the b-axis (Fig. \ref{Temperature}a). The signature of the phase transition is a peak in the dielectric constant near H = 3.8 T. The evolution of the peaks in the dielectric constant (Fig. \ref{Temperature}a) and the magneto-electric current (Fig. \ref{Temperature}b) with temperature are plotted in the phase diagram in Fig. \ref{PhaseDiagram} along with previous neutron scattering and specific heat results \cite{Kenzelmann04,Chen07}. The excellent agreement between the region of long-range magnetic order (AFM B) and electric polarization in Fig. \ref{PhaseDiagram} are evidence for the coexistence and intimate coupling between the electric and magnetic properties of CDC.

Our measurements show that at the upper boundary of the H-T phase diagram in Fig. \ref{PhaseDiagram}, magnetic order and electric polarization are suppressed simultaneously in a continuous phase transition as a function of magnetic field. This transition is not driven by temperature fluctuations, as it occurs at very low temperatures where temperature fluctuations are basically absent. Instead, the transition is the result of magnetic field-tuned quantum fluctuations that are associated with a quantum phase transition at H = 3.8 T \cite{Chen07} where both a magnetic and electric order parameter are critical. This suppression occurs due to a competition between the long-range antiferromagnetic couplings between Cu ions and the effective staggered magnetic fields created by spin-orbit couplings. The resulting quantum fluctuations create a continuous quantum phase transition that is magnetic field-induced. The presence of a peak in the dielectric constant confirms that the electric polarization undergoes an actual phase transition whose order parameter is coupled to that of the magnetic phase transition. This identifies this transition as a magneto-electric quantum phase transition. 

The significance of the observed magnetoelectric quantum phase transition goes well beyond the field of magnetoelectrics and multiferroics. Quantum critical points have been well-investigated for cases where one order parameters is critical, but multi-order quantum phase transitions are only poorly investigated at best. The simultaneity and coupling of the electric and magnetic phase transitions add a new dimension to the field of quantum phase transitions. Investigations of the magnetic component of this quantum critical point using specific heat \cite{Chen07} have shown critical exponents approaching those of a 3-dimensional Ising magnet with effective dimensionality D = d + z = 4, where d = 3 is the spatial dimension and z = 1 is the dynamical exponent. This is contrast to an ordinary thermal Ising phase transition with D = d = 3.

In summary, we demonstrate that magnetoelectric interactions mediated via large molecules produce magnetically-induced ferroelectricity. The ferroelectric polarization can be switched by short magnetic field pulses in an unusual hysteretic fashion. Our study opens the road towards molecule-based designer magnets based on a wider range of organic ligands with which desired magnetoelectric properties can be fine-tuned. Finally, the observation of a magnetoelectric quantum phase transition illustrates the scope of novel physics to be studied in organic magnets.

\begin{acknowledgements}

Work at the National High Magnetic Field Laboratory was supported by the U.S. National Science Foundation through Cooperative Grant No. DMR901624, the State of Florida, and the U.S. Department of Energy. Work at Johns Hopkins University was supported by the National Science Foundation through Grant No. DMR-0306940. 

\end{acknowledgements}


\begin{thebibliography}{10}

\bibitem{Fiebig05}
M. Fiebig, J. Phys. D {\bf 38},  R123  (2005).

\bibitem{Khomskii06}
D.~I. Khomskii, J. Magn. Magn. Mater {\bf 306},  1  (2006).

\bibitem{Eerenstein06}
W. Eerenstein, N.~D. Mathur, and J.~F. Scott, Nature {\bf 442},  759  (2006).

\bibitem{Hill00}
N. Hill, J. Phys. Chem. B {\bf 104},  6694  (2000).

\bibitem{Kimura03}
T. Kimura, T. Goto, H. Shintani, K. Ishizaka, T. Arima, and Y. Tokura, Nature
  {\bf 426},  55  (2003).

\bibitem{Goto04}
T. Goto, T. Kimura, G. Lawes, A.~P. Ramirez, and Y. Tokura, Phys. Rev. Lett.
  {\bf 92},  257201  (2004).

\bibitem{Hur04}
N. Hur, S. Park, P.~A. Sharma, J.~S. Ahn, S. Guha, and S.-W. Cheong, Nature
  {\bf 429},  392  (2004).

\bibitem{Lawes05}
G. Lawes, A.~B. Harris, T. Kimura, N. Rogado, R.~J. Cava, A. Aharony, O.
  Entin-Wolhman, T. Yildirim, M. Kenzelmann, C. Broholm, and A.~P. Ramirez,
  Phys. Rev. Lett. {\bf 95},  087205  (2005).

\bibitem{Katsura05}
H. Katsura, N. Nagaosa, and A.~V. Balatasky, Phys. Rev. Lett. {\bf 95},  057205
   (2005).

\bibitem{Cheong07}
S. Cheong and M. Mostovoy, Nature Materials {\bf 6},  13  (2007).

\bibitem{Kenzelmann07}
M. Kenzelmann, G. Lawes, A.~B. Harris, G. Gasparovic, C. Broholm, A.~P.
  Ramirez, G.~A. Jorge, M. Jaime, S. Park, Q. Huang, A.~Y. Shapiro, and L.~A.
  Demianets, Phys. Rev. Lett. {\bf 98},  267205  (2007).

\bibitem{Arima07}
T. Arima, J. Phys. Soc. Japan {\bf 76},  073702  (2007).

\bibitem{Kimura07}
T. Kimura, Annu. Rev. Mater. Res. {\bf 37},  387  (2007).

\bibitem{Cui06}
H.-B. Cui, Z. Wang, K. Takahashi, Y. Okano, H. Kobayashi, and A. Kobayashi, J.
  Am. Chem. Soc. {\bf 128},  15074  (2006).

\bibitem{Ye08}
Q. Ye, D.-W. Fu, H. Tian, R.-G. Xiong, P.~W.~H. Chan, and S.~D. Huang, Inorg.
  Chem. {\bf 47},  772  (2008).

\bibitem{Scott08}
J.~F. Scott, J. Phys.: Condens. Matter {\bf 20},  021001  (2008).

\bibitem{Willet70}
R.~D. Willet and K. Chang, Inorganic Chem. Act. {\bf 4},  447  (1970).

\bibitem{Kenzelmann04}
M. Kenzelmann, Y. Chen, C. Broholm, D.~H. Reich, and Y. Qiu, Phys. Rev. Lett.
  {\bf 93},  017204  (2004).

\bibitem{Chen07}
Y. Chen, M.~B. Stone, M. Kenzelmann, C.~D. Batista, D.~H. Reich, and C.
  Broholm, Phys. Rev. B {\bf 75},  214409  (2007).

\bibitem{Zapf08}
V.~S. Zapf, V.~F. Correa, P. Sengupta, C.~D. Batista, M. Tsukamoto, N.
  Kawashima, P. Egan, C. Pantea, A. Migliori, J.~B. Betts, M. Jaime, and A.
  Paduan-Filho, Phys. Rev. B {\bf 77},  R020404  (2008).

\bibitem{Schmiedeshoff06}
G.~M. Schmiedeshoff, A.~W. Lounsbury, D.~J. Luna, S.~J. Tracy, A.~J. Schramm,
  S.~W. Tozer, V.~F. Correa, S.~T. Hannahs, T.~P. Murphy, E.~C. Palm, A.~H.
  Lacerda, S.~L. Bud'ko, P.~C. Canfield, J.~L. Smith, J.~C. Lashley, and J.~C.
  Cooley, Rev. Sci. Inst. {\bf 77},  123907  (2006).

\bibitem{Kornev00}
I. Kornev, M. Bichurin, J.-P. Rivera, S. Gentil, H. Schmid, A.~G.~M. Jansen, ,
  and P. Wyder, Phys. Rev. B {\bf 62},  12247  (2000).

\end{thebibliography}
\end{document}